\documentclass[twocolumn,journal]{IEEEtran}
\IEEEoverridecommandlockouts
\usepackage{tabto}
\usepackage{algorithm}
\usepackage{algpseudocode}
\usepackage{paralist}
\usepackage{cite}
\usepackage{amsmath,amssymb,amsfonts}

\usepackage{graphicx}
\usepackage{textcomp}
\usepackage{xcolor}
\usepackage{hyperref}
\usepackage{float}
\usepackage{enumitem}
\usepackage{amsmath}
\usepackage{placeins}

\usepackage{hyperref}

\newcommand{\nix}[1]{}

\begin{document}

\title{\LARGE{Acute Lymphoblastic Leukemia Diagnosis Employing YOLOv11, YOLOv8,  ResNet50, and Inception-ResNet-v2 Deep Learning Models}}

\author{\IEEEauthorblockN{Alaa Awad$^\S$  ~~~ and ~~ Salah A. Aly$^\dag$$^\ddag$\\}
\IEEEauthorblockA{\noindent $^\S$CS \& IT Department, E-Japan University of Science and Tech., Alexandria, Egypt \\
\noindent $^\dag$Faculty of Computing and Data Science, Badya University, Giza, Egypt\\
\noindent $^\ddag$CS \& Math Section, Faculty of Science, Fayoum University, Fayoum, Egypt\\
}
}

\maketitle

\begin{abstract}
Thousands of individuals succumb annually to leukemia alone. As artificial intelligence-driven technologies continue to evolve and advance, the question of their applicability and reliability remains unresolved. This study aims to utilize image processing and deep learning methodologies to achieve state-of-the-art results for the detection of Acute Lymphoblastic Leukemia (ALL) using data that best represents real-world scenarios. ALL is one of several types of blood cancer, and it is an aggressive form of leukemia. In this investigation, we examine the most recent advancements in ALL detection, as well as the latest iteration of the YOLO series and its performance. We address the question of whether white blood cells are malignant or benign. Additionally, the proposed models can identify different ALL stages, including early stages. Furthermore, these models can detect hematogones despite their frequent misclassification as ALL. By utilizing advanced deep learning models, namely, YOLOv8, YOLOv11, ResNet50 and Inception-ResNet-v2, the study achieves accuracy rates as high as 99.7\%, demonstrating the effectiveness of these algorithms across multiple datasets and various real-world situations.

\end{abstract}


\section{Introduction}

Cancer is one of the most lethal diseases known to the human race, and it is embodied in several forms, like leukemia. The incidence rate of leukemia across the globe is estimated at 487,294 with a mortality number of 305,405 annually, according to the International Agency for Research on Cancer, which is part of the World Health Organization (WHO)~\cite{CancerToday}. These statistics highlight the urgency behind the need to establish a better understanding of leukemia and provide effective health services to people affected by this disease.

Leukemia, also known as blood cancer, is one of many types of cancer that originates from the blood and bone marrow. This issue arises from the abnormal and rapid production of white blood cells. Leukemia can be categorized as chronic or acute in relation to the disease’s progression speed and into lymphocytic or myelogenous based on the type of cells it metamorphoses to after growth that can be narrowed to white lymphocyte blood cells or myelogenous, which in turn could be of three types, namely, red blood cells, white blood cells, or platelets~\cite{who2023}.

Nowadays, we live in a technology-themed era in which computer science is exploited to mimic human intelligence for more accurate and faster decision-making. Therefore, many researchers have tried to tackle the leukemia detection problem through artificial intelligence using different deep learning methodologies, such as MobileNetV2~\cite{Sandler2019}, attention mechanism ~\cite{Vaswani2023} and YOLO ~\cite{Redmon2016}. Numerous datasets were used in different studies, like the ALL image dataset~\cite{Ghaderzadeh2021} and the C-NMC 2019 dataset ~\cite{Mourya2019}.

Most of the research done relies on single-cell datasets to train the AI models; however, in a practical environment, the model should expect to be exposed to multi-cell images and still perform accurately. The purpose of this paper is to overcome this gap in research by exposing the proposed model to multi-cell samples.

\begin{figure}[H]
\centerline{\includegraphics[width=8cm, height=5cm]{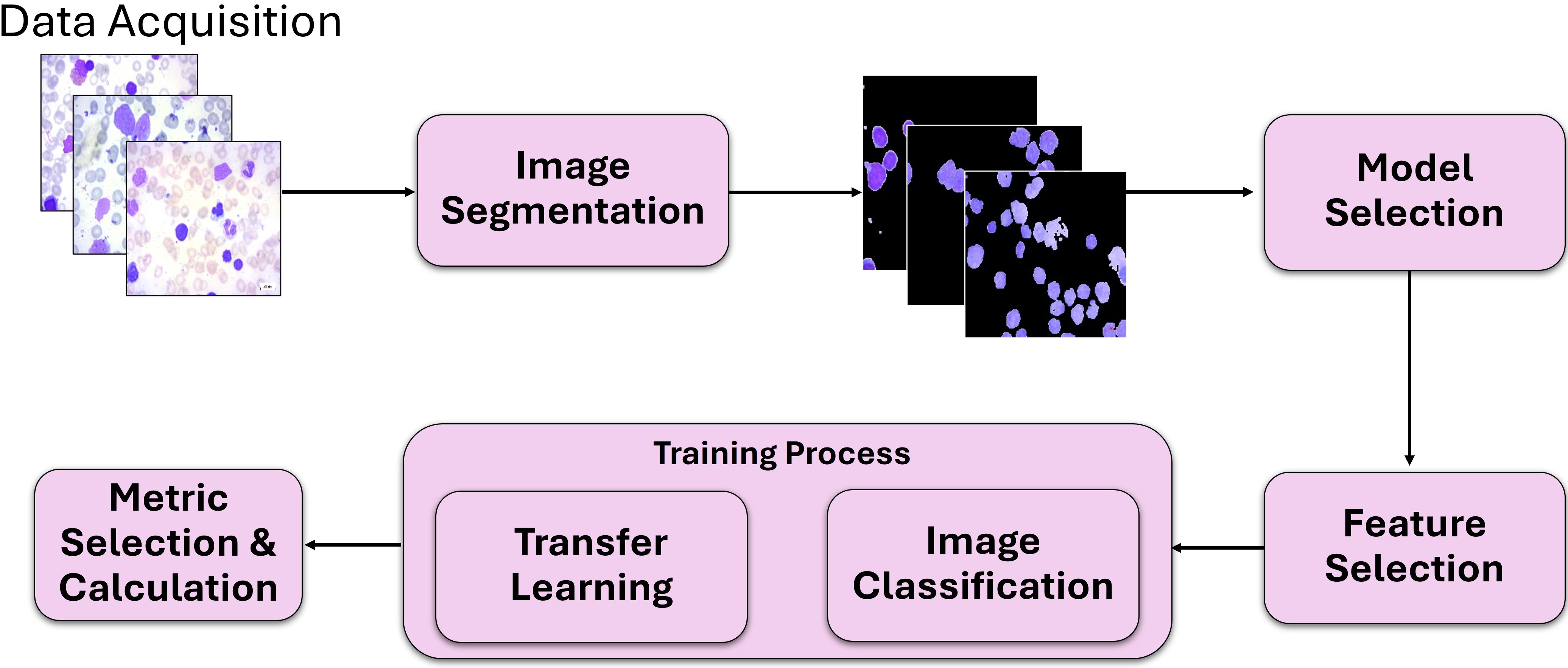}}
\caption{Workflow Diagram}
\label{fig:workflow}
\end{figure}

This study employs image processing techniques, such as segmentation, to prepare the dataset. Additionally, it utilizes transfer learning and fine-tuning methodologies on YOLOv11~\cite{YOLOv11_2024}, YOLOv8~\cite{Varghese2024}, and ResNet50~\cite{He2015}, achieving results ranging from 97\% to 99\%, see our initial results~\cite{awad2024}.

The contributions of this study can be summarized as follows:
\begin{enumerate}
  \item Up to our knowledge, this is the first research conducted to exploit YOLOv11 in blood cancer detection.
  \item The integration of two datasets enhances the model's generalization across diverse samples.
  \item The crucial question of whether white blood cells are malignant or benign is addressed
  \item The models demonstrate the capability to detect hematogones cases, despite their frequent misclassification as ALL.
  \item A comparative analysis of our findings with previous related studies is provided.
\end{enumerate}

The structure of the paper is as follows: The dataset and data collection is presented in section \ref{sec:dataset}. Section \ref{sec:Implementation} goes through the methodologies and several deep learning models used. The performance metrics are discussed in section \ref{sec:metrics}, and the results of YOLOv11, YOLOv8 and ResNet50 are found in sections \ref{sec:yolov11_results}, \ref{sec:yolov8s_results} and \ref{sec:resnet50_results}, respectively. Section \ref{sec:relatedwork} discusses the related work done in this field. Finally, our results are compared with other works in section \ref{sec:resultscomp}, followed by the conclusion in section \ref{sec:conclusion}.

\section{Datasets Description and Data Collections}\label{sec:dataset}
Available datasets are divided into two types: single-cell and multi-cell datasets. Single-cell datasets typically contain images with a single white blood cell per image, whereas multi-cell datasets depict multiple cells within each sample. Since multi-cell datasets better represent real-life scenarios when working with blood cells, we chose to focus on them. The two datasets selected for this study are the Acute Lymphoblastic Leukemia (ALL) image dataset from Kaggle~\cite{Ghaderzadeh2021} and ALL-IDB1~\cite{Genovese2023}, both of which contain multiple white blood cells per sample. We opted for these two datasets because most of the others are single-celled or dedicated to classifying the subtypes of malignant cancer solely.

On one hand, Table~\ref{tab:all_kaggle} summarizes the statistics of the ALL image dataset, which contains 3,256 images in total, divided into four categories: Benign, Early, Pre, and Pro. The benign class includes Hematogones, a condition where lymphoid cells accumulate in a pattern similar to ALL but are non-cancerous and generally harmless. The dataset consists of 504 benign images and 2,752 malignant cells, further categorized into 985 early-stage samples, 963 pre-stage samples, and 804 pro-phase samples.

\begin{table}[htbp]\caption{Sample distribution per class for ALL image dataset }
\label{tab:all_kaggle}
\centering
\begin{tabular}{|c|c|c|}
\hline
\textbf{Class} & \textbf{Samples Per Class} & \textbf{Percentage}\\
\hline
Benign & 504  & 15.5\% \\
Early & 985 & 30.2\%\\
Pre & 963 & 29.6\%\\
Pro & 804 & 24.7\% \\
\hline
\multicolumn{1}{|c|}{\textbf{Total}} & \textbf{3256} &100\% \\
\hline
\end{tabular}

\end{table}

On the other hand, Table~\ref{tab:all_idb1} refers to the ALL-IDB1 dataset, which contains microscopic images. It includes a total of 108 images, divided into 59 normal blood samples and 49 cancerous ones. This balance between normal and cancerous samples is crucial for the model to effectively learn the distinguishing features of ALL cells.

\begin{table}[htbp]\caption{Sample distribution per class for ALL-IDB1 dataset }
\label{tab:all_idb1}
\centering
\begin{tabular}{|c|c|c|}
\hline
\textbf{Class} & \textbf{Samples Per Class}  & \textbf{Percentage}\\
\hline
Normal & 59 & 54.6\%\\
Cancer & 49 & 45.4\% \\
\hline
\multicolumn{1}{|c|}{\textbf{Total}} & \textbf{108} & 100\% \\
\hline
\end{tabular}

\end{table}

We aim to expose the models to diverse datasets and various states of the blast cells for more practical and precise classification. Thus, we merged the normal cells from the ALL-IDB1 dataset with the benign cells from the ALL image dataset under one category called Normal. In addition, we combined the samples from the Early, Pre and Pro classes from the ALL image dataset with the Cancer class samples from the ALL-IDB1 dataset into one class called Cancer. Eventually, the final dataset we will use to train the models will consist of only two classes, Normal and Cancer. Table~\ref{tab:final_db} below clarifies how the integration between the two datasets was done to produce the two final classes in order to ensure that the models would be able to differentiate between healthy and malignant cells even in early stages.
\begin{table}[htbp]
\centering
\caption{Sample distribution per class for the final merged dataset}
\begin{tabular}{|p{1cm}|p{3.75cm}|p{0.8cm}|p{1.2cm}|}
\hline
\textbf{Class} & \textbf{Samples Per Class} & \textbf{Total}  & \textbf{Percentage}\\
\hline
Normal & Normal(ALL-IDB1):59, Benign(ALL Image Dataset):504& 563 & 16.7\%\\
\hline
Cancer & Cancer(ALL-IDB1):49, Early(ALL Image Dataset): 985, Pre(ALL Image Dataset):963, Pro(ALL Image Dataset):804 & 2801 & 83.3\% \\
\hline
\end{tabular}
\label{tab:final_db}
\end{table}

\section{Models and Methodologies}\label{sec:Implementation}
The implementation is divided into several phases, as illustrated in Fig.~\ref{fig:workflow}. The first phase involved data preparation, where image segmentation techniques were applied to isolate the relevant elements. Additionally, image rescaling was done to ensure that each model receives the right shape for the input layer while data augmentation was utilized to assist with the data variety and enhancing the convergence. Next, the pretrained YOLOv11, YOLOv8, ResNet50 and Inception-ResNet-v2 models were loaded, and the last 10 layers were unfrozen for two of them as illustrated. In the final phase, the models are trained to fine-tune the pretrained weights for the specific task at hand.\\

\begin{figure}[h]
\centerline{\includegraphics[width=8cm, height=5cm]{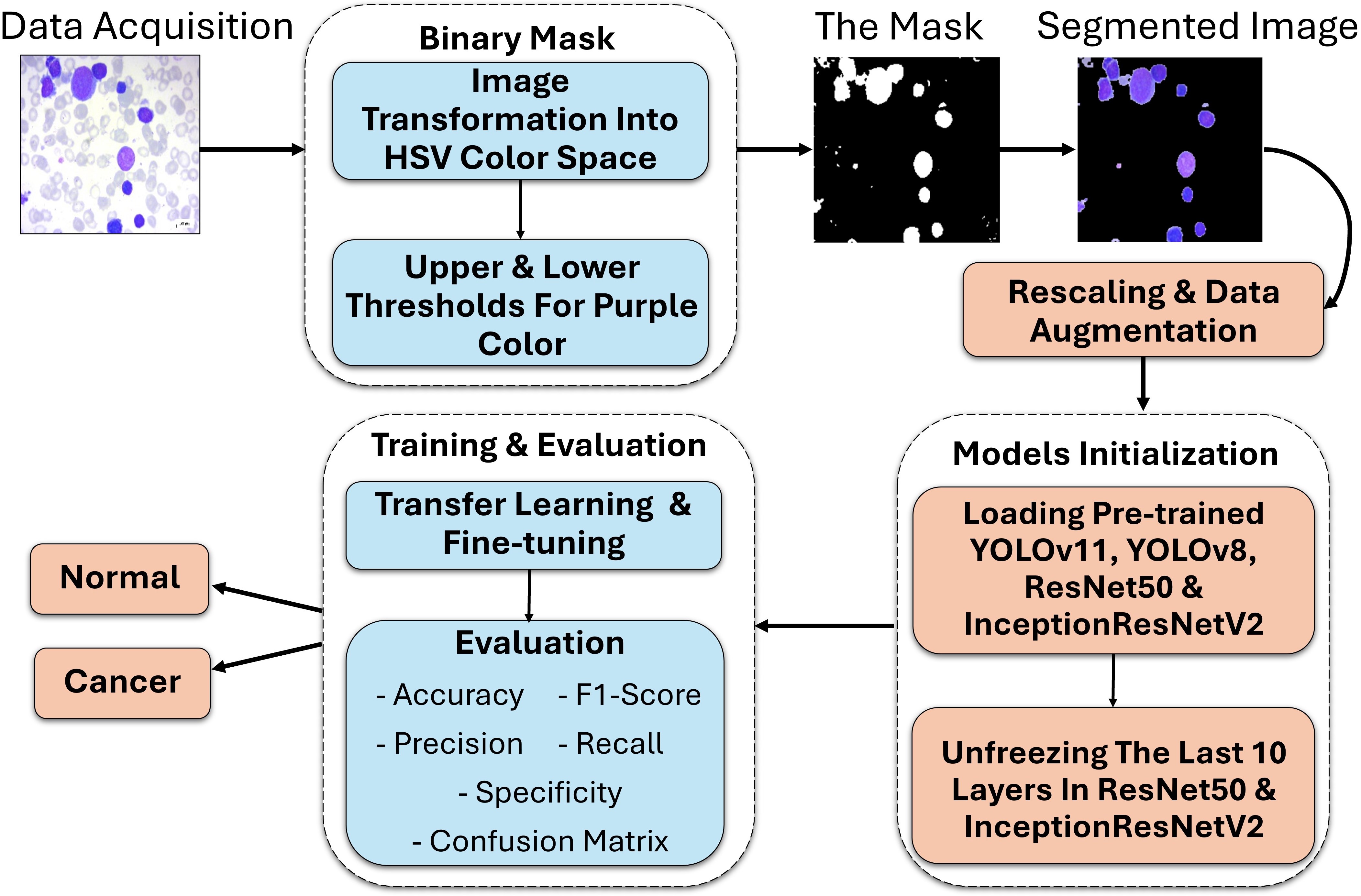}}
\caption{The implementation process}
\label{fig:imp}
\end{figure}

\subsection{Dataset Preparation}
The dataset was preprocessed using image processing techniques to remove redundant elements and improve the models' performance. First, unnecessary elements, such as varying backgrounds and unrelated blood components like platelets, were removed from the images. This step was essential to enable the model to focus on the white blood cells, which are the main area of concern, and reduce potential confusion. To achieve this, image segmentation techniques were applied using OpenCV. The images were converted to the HSV color space (hue, saturation, value), and upper and lower thresholds were set for the purple hue of the white blood cells to create a binary mask. This mask was then applied to the original images ~\ref{fig:orig}, allowing the segmentation of the white blood cells ~\ref{fig:segm}, as illustrated in Fig. ~\ref{fig:Dataset} below.
\begin{figure}[htbp]
\centering
    \includegraphics[width=8cm, height=4cm]{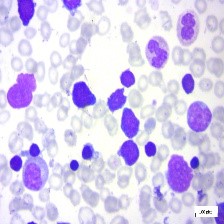}\\
    (a) Original data sample
\hfill
    \includegraphics[width=8cm, height=4cm]{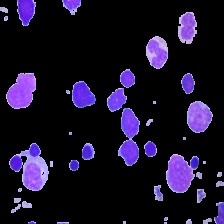}
    (b)Segmented data sample
    \label{fig:segm}
\caption{Data samples before and after image segmentation. (a) Before segmentation and (b) After segmentation.}
\label{fig:Dataset}
\end{figure}

To introduce more robustness and reduce potential overfitting, the data was augmented using random flipping, rotation, and zoom for ResNet50 and InceptionResNetV2. Meanwhile, the training configuration for the YOLO models incorporated several augmentation techniques to enhance robustness and generalization. First, the augmentation parameter was enabled to allow training data augmentation. Mosaic augmentation, which combines four images into one during training, was applied with a probability of 1.0, meaning it was used 100\% of the time. We allowed images to be randomly rotated by up to 45 degrees and added a 50\% chance of flipping them horizontally. Moreover, we applied a scaling factor of 0.5, meaning images could be resized or zoomed in/out by up to 50\%, either enlarging or shrinking them. Finally, the dataset was divided into three sets—training, validation, and testing—with respective ratios of 70\%, 15\%, and 15\%.
\subsection{Image Classification}

The detection of blast cells can be done in numerous ways; one of them is image classification. Countless deep learning architectures have been introduced to support this task. Convolutional Neural Networks (CNNs) are the most common architectures used when working with image and video datasets. Many models have been built based on CNNs such as VGG~\cite{Simonyan2015}, ResNet~\cite{He2015}, AlexNet~\cite{Krizhevsky2012} and GoogleNet (Inception)~\cite{Szegedy2014}. In this paper, the light will be shed on two versions of YOLO, which are 8 and 11, in addition to ResNet50. To train and optimize the models' performance, we used the following two techniques:

\textbf{Transfer Learning:} It is a method adopted in the deep learning field. It aims to reduce the training time and cost for sophisticated models and reuse the massive number of trained weights. Additionally, it is used to overcome the shortage in the available datasets needed to train the models. The core idea behind transfer learning is to allow the model to transfer its knowledge from one task to another. This mechanism works by training a large deep learning model, then exploiting the trained parameters in another task, so the model would not have to learn things from scratch; consequently, the model will not have to reinitialize random weights. Instead, it will build on the prior knowledge to customize to fit with the new task~\cite{Iman2023}.

\textbf{Fine-tuning:} It is a transfer learning approach where the some of the architecture's layer are frozen and the others are unfrozen for further training on a certain task's data. Usually, the layers the unfrozen layers are the last few layers in the model, so the model can adapt to the new task while using the general knowledge it was trained on at the beginning~\cite{Iman2023}.

\textbf{YOLOv8}: One of the recent advancements in computer vision is YOLO~\cite{Redmon2016} which is based on CNNs. YOLOv8~\cite{Varghese2024} is one of the recent released versions in the series of YOLO object detection algorithms. The architecture of this deep learning model consists of two main components: the backbone network and the detection head. The backbone network, based on EfficientNet~\cite{Mehla2023}, extracts rich, multi-scale features from the input image, while the detection head, built using NAS-FPN, merges these features to generate high-quality predictions for object detection. Several new features enhance YOLOv8’s performance. These include the Focal Loss function, which focuses on hard-to-classify examples, reducing the impact of easier cases. Mixup is a new data augmentation technique that blends images and their labels to improve generalization and robustness. Additionally, a new evaluation metric, Average Precision Across Scales (APAS), evaluates object detection accuracy across different object sizes, providing a more comprehensive performance measure than standard metrics.

\textbf{YOLOv11}: YOLOv11 ~\cite{YOLOv11_2024} is the latest release of YOLO by Ultralytics to continue the legacy of the YOLO series. YOLOv11 introduces several key advancements over previous versions in the YOLO series. It features an enhanced backbone and neck architecture for improved feature extraction, boosting accuracy in object detection tasks. The model is optimized for both speed and efficiency, maintaining high performance with fewer parameters compared to earlier versions. YOLOv11 supports a wide range of computer vision tasks, including object detection, image classification, segmentation, and pose estimation, and is adaptable for deployment across diverse platforms, including edge devices and cloud environments.

\textbf{ResNet50}: ResNet~\cite{He2015} is a deep convolutional neural network architecture that has 50 layers which was inspired by VGG nets~\cite{Simonyan2015}, and its network design consists of the plain network and Residual network. It was developed to address the vanishing gradient problem common in very deep networks by introducing residual blocks—layers that skip connections to allow gradients to flow more effectively during training. These residual connections enable the model to focus on learning new features without overwriting already-learned patterns. Moreover, this type of skip link has the advantage of allowing regularization to bypass any layer that impairs architecture performance. ResNet50’s architecture has demonstrated remarkable effectiveness, especially in image classification tasks, and has become a foundation for more advanced models in computer vision.

\textbf{Inception-ResNet}: In the residual versions of the Inception networks, Inception blocks are simplified and optimized to reduce computational costs. Each block is followed by a filter-expansion layer, a 1x1 convolution without activation, which adjusts the filter bank's depth to match the input dimensions after each block. This approach compensates for any reduction in dimensionality introduced by the Inception blocks. Two main versions, Inception-ResNet-v1 and Inception-ResNet-v2, were developed; the former aligns with Inception-v3’s computational demands, while the latter matches Inception-v4’s.
A key design choice in Inception-ResNet was the selective use of batch normalization, applied only to traditional layers and not to summation points, which helped manage GPU memory usage and allowed for more Inception blocks. This decision aimed to keep the models trainable on a single GPU, with the goal of eventually revisiting this compromise as computing efficiency improves~\cite{Szegedy2016}.

\textbf{Methodology}:
First, transfer learning was leveraged in this implementation, where we imported a pretrained YOLOv8 model after installing the Ultralytics package, then trained it on our customized segmented dataset. We experimented with different optimizers and hyperparameters; however, the final result is based on 50 epochs of training using SGD optimizer and a learning rate of 0.001. The batch and image sizes were set to 8 and 224, respectively.

Second, YOLOv11, which is the latest version of the series, was trained on our custom data as well. In this step, we experimented with both the nano and small versions of the model to observe the performance on the different complexity levels of YOLOv11's architecture, eventually deciding which one is more suitable for this problem. The best results were chosen based on training the model with 50 epochs, SGD optimizer, 0.001 learning rate, and 32 batch size.

Besides deploying different YOLO versions, ResNet50 and Inception-ResNet-v2 were also exploited. The segmented dataset was loaded and prepared using image\_dataset\_from\_directory from TensorFlow, with the batch size set to 8 for better generalization. The images were resized to (224, 224, 3) and (299, 299, 3) to match the input layers of ResNet50 and Inception-ResNet-v2, respectively. For both models, pre-trained versions with ImageNet weights were imported from Keras. Although the top layers were frozen, the last 10 layers of the architecture were unfrozen for fine-tuning to better customize the models for the current task. Additionally, the initial learning rate was set to 0.001, and a ReduceLROnPlateau scheduler was added to lower the learning rate as training progressed, helping to better control convergence. Many experiments were done with larger batch sizes and different number of unfrozen layers but they yielded in lower accuracies, more unstable performances or overfitting. Therefore, the parameters mentioned led to the best performance.

\begin{figure}\label{alg1}
\caption{YOLOv8, YOLOv11, ResNet50 and Inception-ResNet-v2 Models Training and Evaluation}
\begin{algorithmic}[1]
\Require ALL-IDB1 and ALL image dataset on CNN models.
\Ensure Accuracy, Loss and saved models.
\State \textbf{Step 1: Model Initialization}
   \State~~-~Load pre-trained YOLOv8, YOLOv11, ResNet50 and Inception-ResNet-v2.
   \State~~-~Unfreeze the last 10 layers in both ResNet50 and Inception-ResNet-v2.
   \State~~-~Initialize  parameters for each model, including learning rate, optimizer, loss and metric.
\State \textbf{Step 2: Data Preprocessing and Analysis}
   \State~~-~Perform image segmentation on ALL-IDB1 and ALL image dataset.
   \State~~-~Rescale and apply data augmentation properly on the datasets.
\State \textbf{Step 3: Model Training and Evaluation}
   \State~~- Train YOLOv8 and YOLOv11 models on the preprocessed data through transfer learning.
   \State~~- Fine-tune ResNet50 and Inception-ResNet-v2 on the preprocessed data.
   \State~~- Validate  performance and tune the parameters using the validation dataset.
   \State~~-  Evaluate  performance of the models on the test dataset.
\State \textbf{Step 4: Model Evaluation}
   \State~~- Calculate the accuracy, Recall, Precision, F1-Score and Specificity.
   \State~~- Generate the training and validation accuracies and losses graphs.
   \State~~- Generate the confusion matrix.
\end{algorithmic}
\end{figure}

\topmargin=-0.7in

\medskip
\section{Performance Metrics}\label{sec:metrics}
Numerous evaluation methods and metrics have been developed to assess different types of tasks in deep learning. In this study, we have used several metrics to illustrate and comprehend the efficiency of the models used. These calculations provided us with a clear explanation for the results obtained from each model during the experimentation phase, which assisted and guided us when tweaking the methodologies for optimal results. For this purpose, accuracy, f1 score, precision, recall, specificity, and confusion matrix were computed.

Accuracy is an overall indicator of how well the model performs, considering the number of correctly identified samples out of all the given samples. It is represented by the sum of true positives and true negatives divided by the total number of examples, which includes True Positive (TP), True Negative (TN), False Positive (FP), and False Negative (FN), as expressed in Equation~\ref{eq:acc}:
\begin{equation}
 Accuracy = \frac{TP+TN}{TP+TN+FP+FN}
 \label{eq:acc}
\end{equation}

Presented in Equation~\ref{eq:prec}, the sample precision which is identified by the ratio of correctly classified instances to the total number of classified instances.
\begin{equation}
 Precision = \frac{TP}{TP+FP}
 \label{eq:prec}
\end{equation}

Recall, or Sensitivity, is calculated as the ratio of correctly identified instances to the total number of instances, as described in Equation~\ref{eq:recall}.
\begin{equation}
 Recall = \frac{TP}{TP+FN}
 \label{eq:recall}
\end{equation}

Another significant metric that contributed to our results is F1-score. It is obtained by calculating the harmonic mean of precision and recall, as illustrated in Equation~\ref{eq:f1}.
\begin{equation}
 F1 Score = 2*\frac{Precision * Recall}{Precision + Recall}
 \label{eq:f1}
\end{equation}

In addition to the previous indicators, we calculated the specificity using the formula in Equation~\ref{eq:spec}.It refers to the proportion of correctly identified negative instances among all actual negative cases. It reflects the model's ability to accurately classify instances from the opposite disease classes.
\begin{equation}
 Specificity = \frac{TN}{TN+FP}
 \label{eq:spec}
\end{equation}

Lastly, we generated the confusion matrix, which provides a comprehensive summary of the model's performance and the essential components required to derive the previously mentioned metrics. This matrix compares the predicted classes of samples to their actual classes by displaying the counts for True Negatives (TN), False Positives (FP), True Positives (TP), and False Negatives (FN). It can be represented as follows~\ref{eq:matrix}:

\begin{equation}
\begin{bmatrix}
\text{TP} & \text{FP} \\
\text{FN} & \text{TN}
\label{eq:matrix}
\end{bmatrix}
\end{equation}
\section{YOLOv11 Performance Results}\label{sec:yolov11_results}
In this section, we evaluate the performance of our various trained models using the metrics mentioned in section~\ref{sec:metrics}. Starting with YOLOv11, as mentioned before, we employed two versions of the model to test different complexities and determine the most appropriate for this problem. Both the nano and small models were trained for 50 epochs, with YOLOv11n achieving 98\% validation accuracy and 97.3\% testing accuracy, while YOLOv11s attained a higher testing accuracy of 98.2\%.

\begin{figure}[htbp]
\centering
\includegraphics[width=0.7\linewidth]{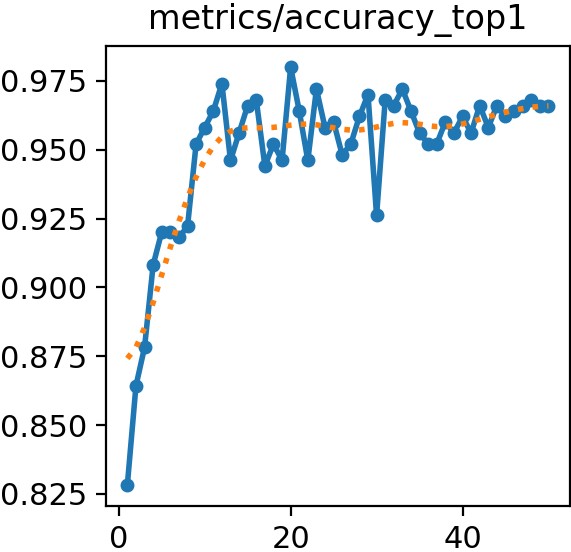}
\caption{YOLOv11n accuracy with SGD}
\label{fig:accuracy_yolov11n_sgd}
\end{figure}

The accuracy graph for the nano version of YOLOv11 shown in Fig.~\ref{fig:accuracy_yolov11n_sgd} demonstrates improvement in the accuracy's progress with some fluctuations at the beginning. These variations decrease gradually as the number of epochs increases until the graph curve becomes more stable. It is important to mention that this evaluation was based on SGD optimizer and a learning rate of 0.001.

\begin{figure}[htbp]
\centering
\includegraphics[width=0.7\linewidth]{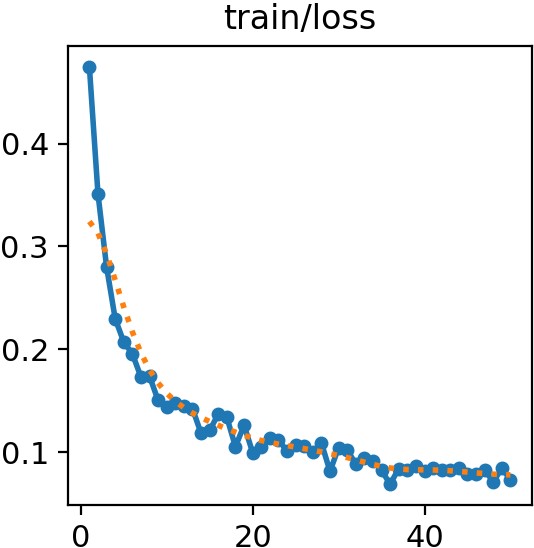}
\caption{YOLOv11n train loss with SGD}
\label{fig:train_loss_yolov11n_sgd}
\end{figure}

\begin{figure}[htbp]
\centering
\includegraphics[width=0.7\linewidth]{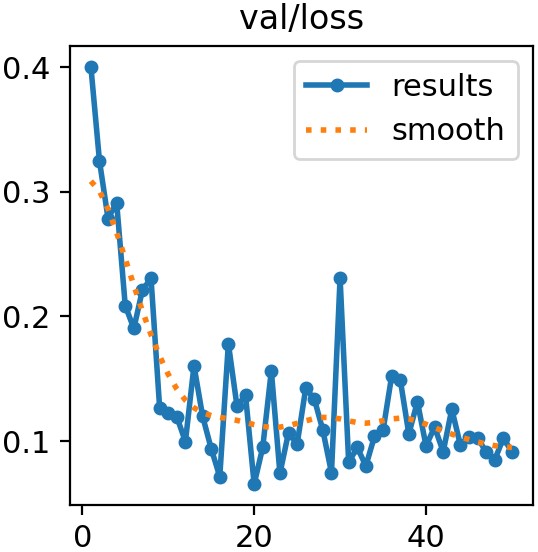}
\caption{YOLOv11n validation loss with SGD}
\label{fig:val_loss_yolov11n_sgd}
\end{figure}

It can be seen that both the training and validation losses were declining as the training advanced in Figures.~\ref{fig:train_loss_yolov11n_sgd} and ~\ref{fig:val_loss_yolov11n_sgd}. The training loss was going down steadily while the validation loss experienced more changes during its decline.

\begin{figure}[htbp]
\centering
\includegraphics[width=\columnwidth]{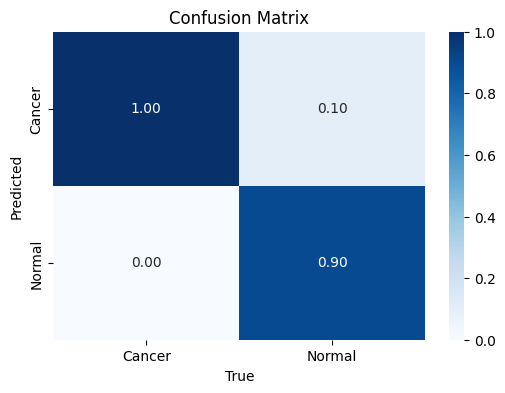}
\caption{Normalized confusion matrix for YOLOv11n with SGD.}
\label{fig:conf_matrix_yolov11n}
\end{figure}

Analyzing the confusion matrix in Fig.~\ref{fig:conf_matrix_yolov11n} provides comprehensive and direct insights about the model's performance because it allows us to identify which classes the model can detect better and when it usually makes poor decisions. Eventually, it helps us decide what to fix and optimize to boost the model's behavior. The matrix for YOLOv11n clarifies how the model obtained its high accuracy and validates the model's efficient overall performance. It seems that the trained model performed significantly well when detecting cancer in all its stages; however, it made some mistakes when identifying the healthy white blood cells, as it mistook 0.10\% for cancerous cells.

\begin{figure}[htbp]
\centering
\includegraphics[width=0.7\linewidth]{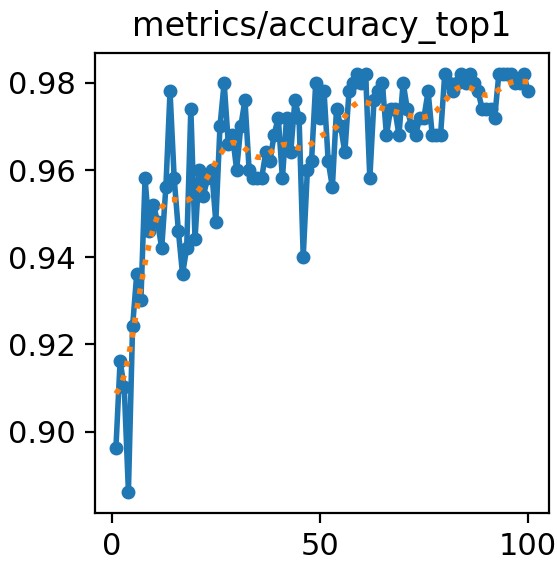}
\caption{YOLOv11n accuracy with AdamW}
\label{fig:accuracy_yolov11n}
\end{figure}

\begin{figure}[htbp]
\centering
\includegraphics[width=0.7\linewidth]{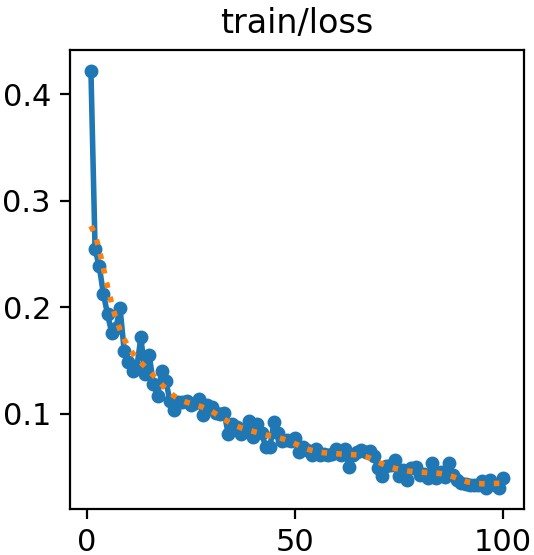}
\caption{YOLOv11n train loss with AdamW}
\label{fig:train_loss_yolov11n}
\end{figure}

\begin{figure}[htbp]
\centering
\includegraphics[width=0.7\linewidth]{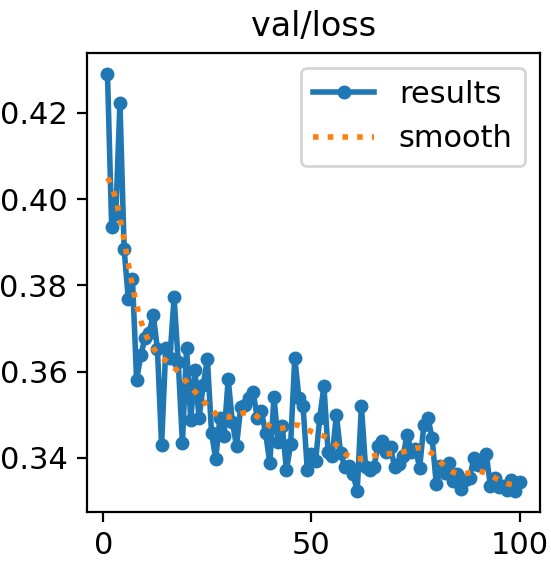}
\caption{YOLOv11n validation loss with AdamW}
\label{fig:val_loss_yolov11n}
\end{figure}

We also conducted experiments with different optimizers and hyperparameters. Figures~\ref{fig:accuracy_yolov11n},~\ref{fig:train_loss_yolov11n}, and~\ref{fig:val_loss_yolov11n} exhibit the model's performance when leveraging AdamW optimizer and the default 0.000714 YOLO learning rate during 100 epochs on batches of size 16. AdamW optimizer shows more unsteady performance than SGD in the accuracy and validation loss graphs. Furthermore, it can be noticed that more epochs lead to slower convergence, where the validation loss stabilizes later after around 60 epochs, with more frequent oscillations even in later epochs. Despite this experiment achieving a validation accuracy of 98.2\%, which is higher than that of SGD optimizer, the model's behavior here shows a greater risk of overfitting. Thus, we decided to choose the YOLOv11n with SGD as the best performance achieved, taking into consideration not only the accuracy but also the stability and efficiency.

\begin{figure}[htbp]
\centering
\includegraphics[width=0.7\linewidth]{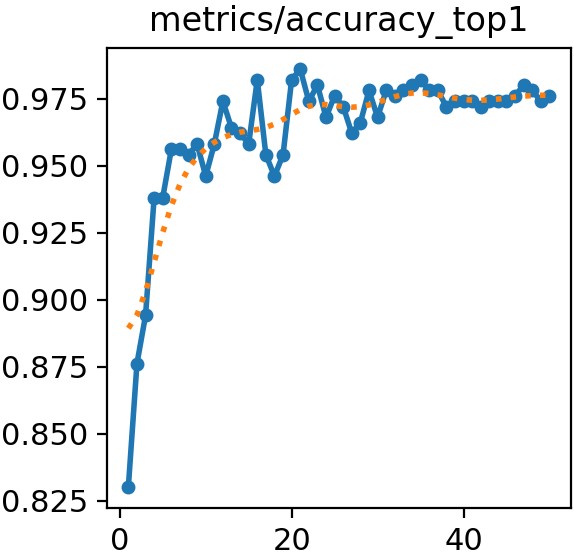}
\caption{YOLOv11s accuracy with SGD}
\label{fig:accuracy_yolov11s_sgd}
\end{figure}

\begin{figure}[htbp]
\centering
\includegraphics[width=0.7\linewidth]{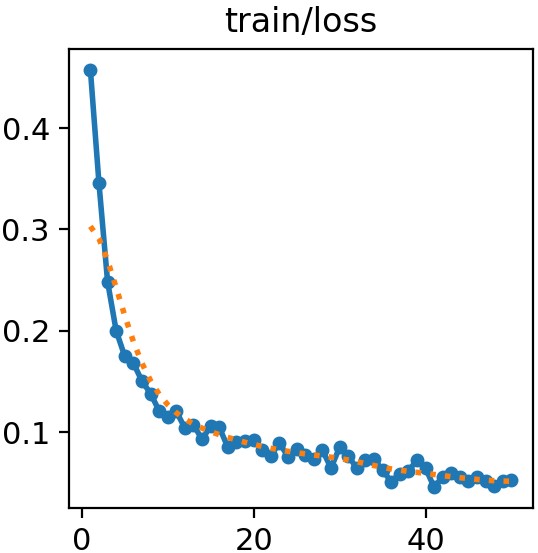}
\caption{YOLOv11s train loss with SGD}
\label{fig:train_loss_yolov11s_sgd}
\end{figure}

\begin{figure}[htbp]
\centering
\includegraphics[width=0.7\linewidth]{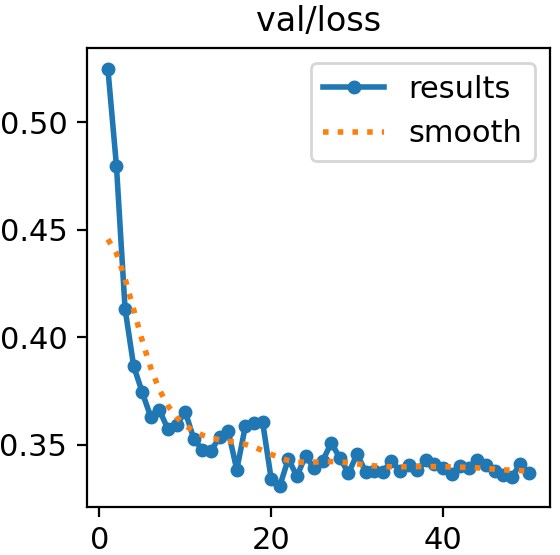}
\caption{YOLOv11s validation loss with SGD}
\label{fig:val_loss_yolov11s_sgd}
\end{figure}

YOLOv11s exhibited a similar but enhanced behavior compared to YOLOv11n. The optimal performance on YOLOv11n was reached with SGD optimizer, 32 batch size, and 0.001 learning rate, and the model was trained for 50 epochs. Fig.~\ref{fig:accuracy_yolov11s_sgd} clarifies how the accuracy value followed the same trend as YOLOv11n, where it rose with some random variations at the beginning indicating that the model was still learning the new data patterns. The graph became more stable by the end of the training process, achieving 98.6\% validation accuracy. In addition, the test accuracy reached 98.2\%, surpassing that of Yolov11n by 0.9. Figures~\ref{fig:train_loss_yolov11s_sgd} and ~\ref{fig:val_loss_yolov11s_sgd} demonstrate the considerable decline in both the training and validation losses.

\begin{figure}[htbp]
\centering
\includegraphics[width=\columnwidth]{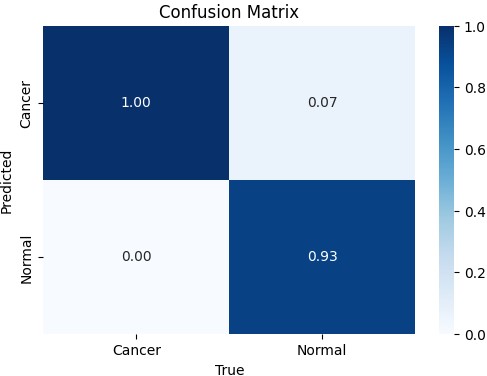}
\caption{Normalized confusion matrix for YOLOv11s with SGD.}
\label{fig:conf_matrix_yolov11s_sgd}
\end{figure}

There was a slight improvement in the confusion matrix as well, which can be illustrated in Fig.~\ref{fig:conf_matrix_yolov11s_sgd}, such that the rate of images of healthy white blood cells misclassified as cancer was less than that of YOLOv11n since it fell from 0.10 to 0.07.

On the other hand, Figures~\ref{fig:accuracy_yolov11s_sgd},~\ref{fig:train_loss_yolov11s_sgd}, and~\ref{fig:val_loss_yolov11s_sgd} visualize the model's performance when trained using the AdamW optimizer. Although it attained a higher validation accuracy of 98.8\%, the training process showed considerable fluctuations, reflecting instability and a possible risk of overfitting. Consequently, we opted for model trained with SGD, as it demonstrated superior generalization and more stable performance. The experiments with YOLOv11 revealed several important insights. Training with the SGD optimizer produced smoother training and validation curves, whereas the AdamW optimizer achieved marginally higher accuracy. Moreover, larger batch sizes contributed to improved accuracy. However, increasing the number of epochs beyond 50 caused a drop in performance, with 100 or more epochs proving detrimental. As a result, 50 epochs were chosen as the optimal configuration.

\begin{figure}[htbp]
\centering
\includegraphics[width=0.7\linewidth]{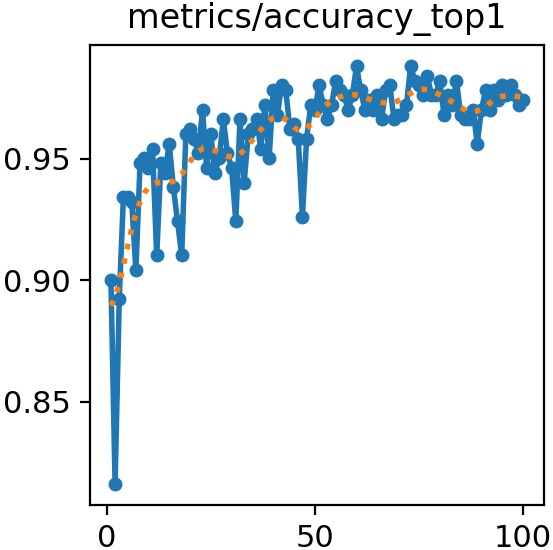}
\caption{YOLOv11s accuracy with AdamW}
\label{fig:accuracy_yolov11s}
\end{figure}
\begin{figure}[htbp]
\centering
\includegraphics[width=0.7\linewidth]{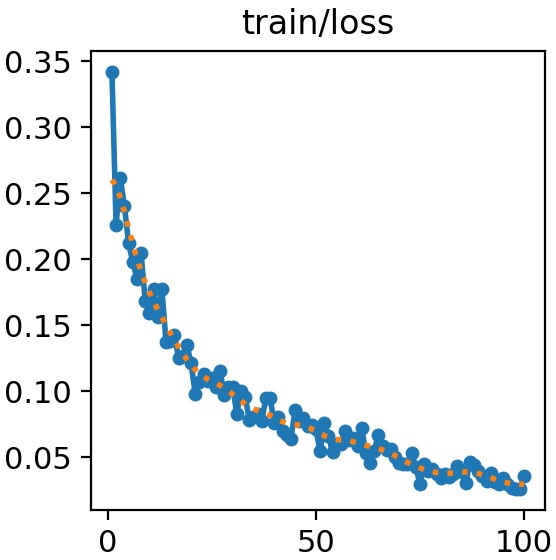}
\caption{YOLOv11s train loss with AdamW}
\label{fig:train_loss_yolov11s}
\end{figure}

\begin{figure}[htbp]
\centering
\includegraphics[width=0.7\linewidth]{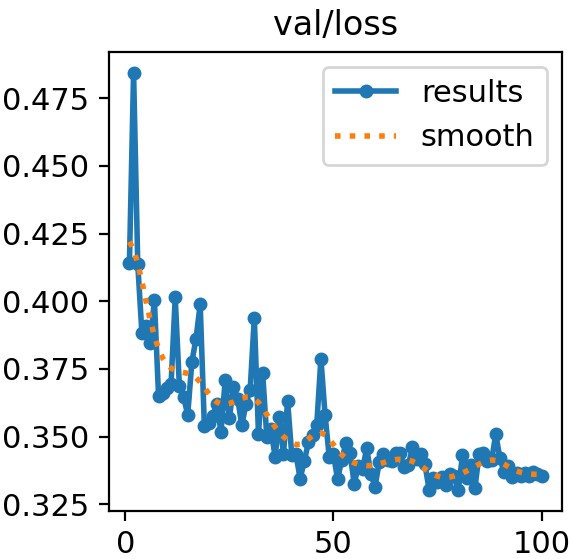}
\caption{YOLOv11s validation loss with AdamW}
\label{fig:val_loss_yolov11s}
\end{figure}

\section{YOLOv8s Performance Results}\label{sec:yolov8s_results}
\begin{figure}[htbp]
\centering
\includegraphics[width=0.7\linewidth]{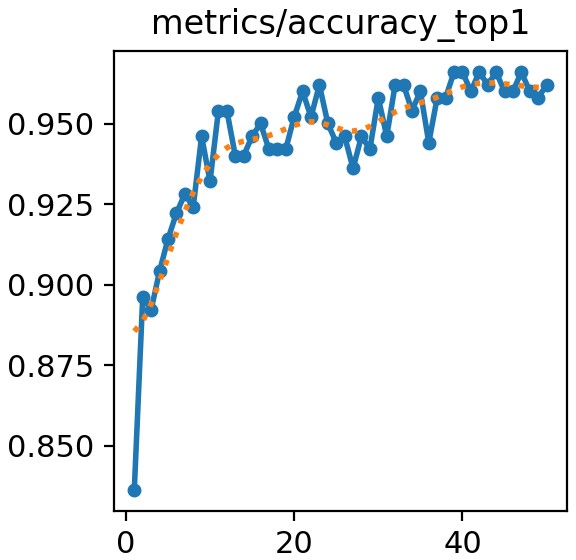}
\caption{YOLOv8s accuracy with SGD}
\label{fig:accuracy_yolov8s}
\end{figure}

\begin{figure}[htbp]
\centering
\includegraphics[width=0.7\linewidth]{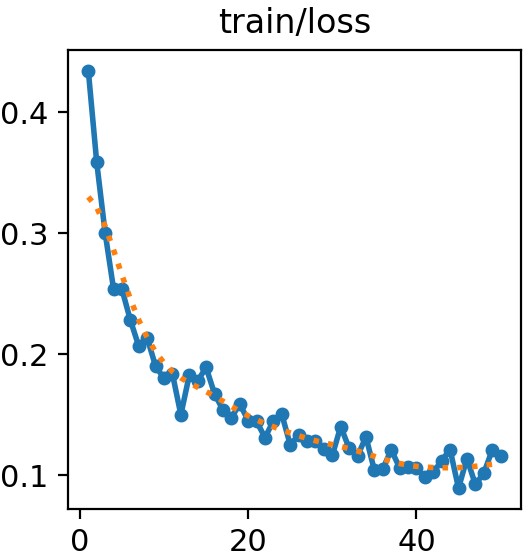}
\caption{YOLOv8s train loss with SGD}
\label{fig:train_loss_yolov8s}
\end{figure}

\begin{figure}[htbp]
\centering
\includegraphics[width=0.7\linewidth]{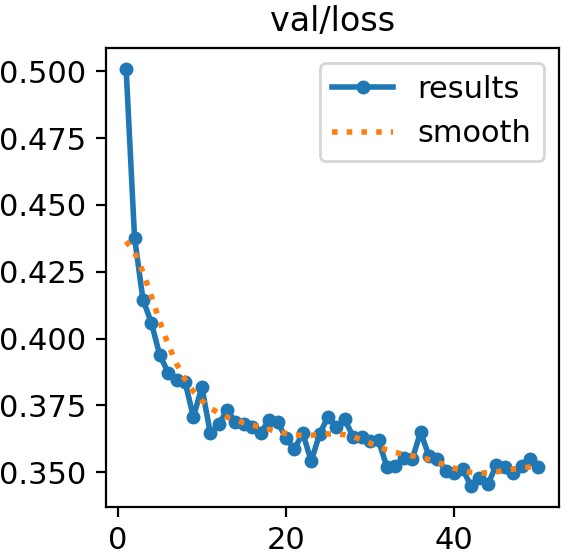}
\caption{YOLOv8s validation loss with SGD}
\label{fig:val_loss_yolov8s}
\end{figure}

The visual representation of YOLOv8 behavior is shown in figures ~\ref{fig:accuracy_yolov8s},~\ref{fig:train_loss_yolov8s}, and ~\ref{fig:val_loss_yolov8s}, in which the accuracy on the validation dataset was 96.6\%, and it peaked at  98\% when evaluated on the testing dataset. Compared to YOLOv11s, YOLOv8’s small version achieves a slightly lower accuracy while it outperforms the nano version. The accuracy, training loss, and validation loss graphs of YOLOv8 presented follow similar patterns as those of YOLOv11. In comparing optimizers, YOLOv8 demonstrated greater stability when using SGD rather than AdamW. For batch size, we experimented with 8, 16, 32, and 64, finding that smaller batch sizes yielded better performance. Consequently, a batch size of 8 was chosen for the final model.

\begin{figure}[htbp]
\centering
\includegraphics[width=\columnwidth]{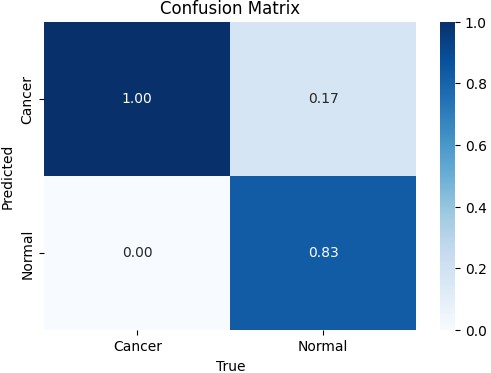}
\caption{Confusion matrix for YOLOv8s with SGD}
\label{fig:conf_matrix_yolov8s}
\end{figure}

The confusion matrix in Fig.~\ref{fig:conf_matrix_yolov8s} illustrates how the number of misclassified normal cells surpasses that of the small and nano versions of YOLOv11; otherwise, the model is performing considerably well and efficiently.

\section{ResNet50 Performance Results}\label{sec:resnet50_results}
\begin{figure}[htbp]
\centering
\includegraphics[width=\linewidth]{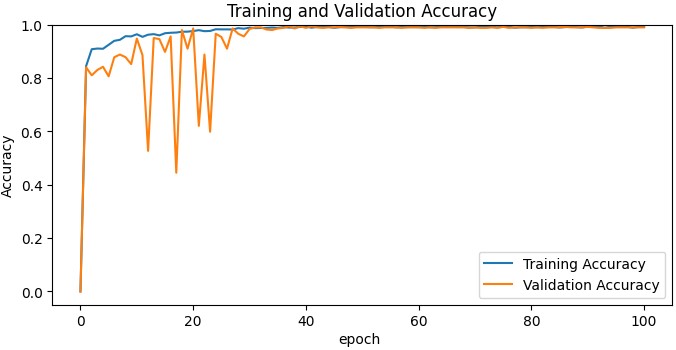}
\caption{Training and Validation Accuracy}
\label{fig:train_val_acc_resnet50}
\end{figure}

\begin{figure}[htbp]
    \includegraphics[width=\linewidth]{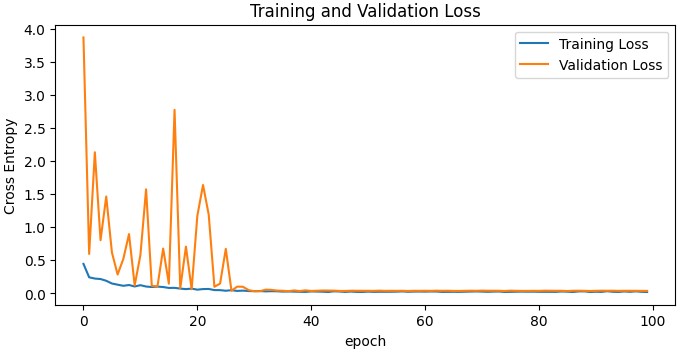}
    \caption{Training and Validation Loss}
    \label{fig:train_val_loss_resnet50}
\end{figure}

In this section, we evaluate ResNet50's behavior and performance metrics. Using fine-tuning, the model’s training accuracy settled at a peak of 99.2\%. Meanwhile, the validation accuracy grew to 99\%, and the test dataset evaluation achieved 99\%. The visualization of the training and validation accuracy curves in Fig.~\ref{fig:train_val_acc_resnet50} highlights the learning enhancement and growth along the epochs. There were some variations at the beginning of the validation curve that were reduced as the training process progressed, and there was almost no gap between the training and validation by the hundredth epoch, which also applies to the training and validation losses in Fig.~\ref{fig:train_val_loss_resnet50}.

\vspace{0.5em}

\begin{figure}[htbp]
\centering
\includegraphics[width=\columnwidth]{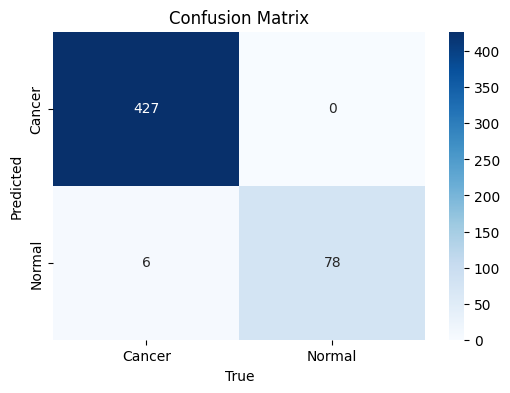}
\caption{Confusion Matrix for ResNet50}
\label{fig:conf_matrix_resnet50}
\end{figure}

On a different note, the confusion matrix of ResNet50 in Fig.~\ref{fig:conf_matrix_resnet50} illustrates that there is a percentage of healthy white blood samples that was misidentified as cancerous. However, it identifies all of the blast cells correctly.

\section{Inception-ResNet-v2 Performance Results}\label{sec:Inception-ResNet-v2_results}
\begin{figure}[htbp]
\centering
\includegraphics[width=\linewidth]{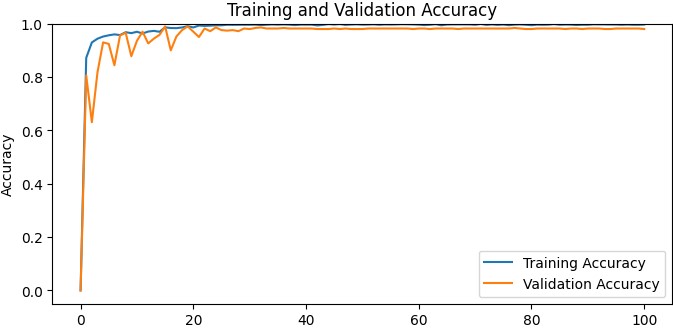}
\caption{Training and Validation Accuracy}
\label{fig:train_val_Inception-ResNet-v2_acc}
\end{figure}

\begin{figure}[htbp]
    \includegraphics[width=\linewidth]{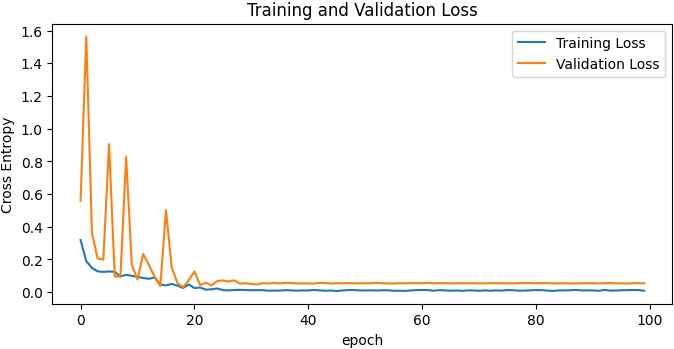}
    \caption{Training and Validation Loss}
    \label{fig:train_val_Inception-ResNet-v2_loss}
\end{figure}

\begin{figure}[htbp]
\centering
\includegraphics[width=\columnwidth]{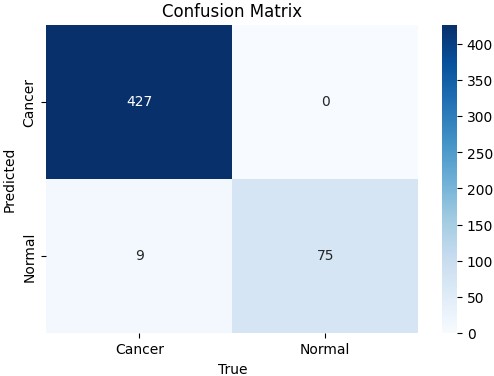}
\caption{Confusion Matrix for Inception-ResNet-v2}
\label{fig:conf_matrix_Inception-ResNet-v2}
\end{figure}

Finally, we analyze the performance of the fine-tuned Inception-ResNet-v2 model, where the model's behavior is represented in figures ~\ref{fig:train_val_Inception-ResNet-v2_acc}, ~\ref{fig:train_val_Inception-ResNet-v2_loss} and ~\ref{fig:conf_matrix_Inception-ResNet-v2}. The training and validation accuracy graph illustrates the training progress across epochs, which was initially unstable for the validation but eventually converged smoothly, recording training, validation, and testing accuracies of 99.7\%, 98\%, and 99.7\%, respectively. Meanwhile, the training and validation losses exhibited a similar trend but in the opposite direction, declining over time with some sharp variations for the validation at the start of the process. It is worth noting that, compared to ResNet50, Inception-ResNet-v2 demonstrated a more stable training process.

\vspace{0.5em}

The confusion matrix of the model is visualized in Fig.~\ref{fig:conf_matrix_Inception-ResNet-v2}, which highlights the model's strong capability in distinguishing between cancerous and normal samples with very few misclassifications.

Table~\ref{tab:performance_metrics} highlights the key findings of this study by evaluating the algorithms on the test dataset using various evaluation metrics. To sum up, it is notable that Inception-ResNet-v2 comes first in both accuracy and specificity among the five models with 99.7\% and 100\% accuracy and specificity, respectively. YOLOv11s comes second in specificity with 93\%, while YOLOv8s comes in last place. It is evident that YOLOv8s attained lower scores than the rest of the tested algorithms for the F1 and precision metrics too. However, YOLOv11n falls at the bottom of the list when considering the accuracy since it has only 97.3\%.

\begin{table}[htbp]\caption{Performance metrics for different models}
\label{tab:performance_metrics}
\centering
\begin{tabular}{|p{1.2cm}|c|c|c|c|c|}
\hline \hline
  &   &   &   &   &   \\
Model & Accuracy & F1 & Precision & Recall & Specificity \\
&   &   &   &   &   \\
\hline \hline
YOLOv11n &  97.3 & 98.8 & 97.9 & 99.8 & 89.5 \\
YOLOv11s & 98.2 & 99.2 & 98.6 & 99.8 & 93 \\
YOLOv8s & 98 & 98 & 96.5 & 99.6 & 82.6 \\
ResNet50 & 99 & 99.3 & 98.6 & 100 & 92.8 \\
Inception-ResNet-v2 & 99.7 & 98.96 & 100 & 97.94 & 100 \\
\hline
\end{tabular}

\end{table}

\section{Related Work}\label{sec:relatedwork}

Extensive research has been conducted on the application of AI for Leukemia detection. This section sheds light on the previous work done on Leukemia detection using different Deep Learning architectures, along with their strengths and weaknesses.

Hosseini et al.~\cite{Hosseini2023} strove to detect B-cell acute lymphoblastic leukemia (B-ALL) cases, including the subtypes, using a deep CNN. After leveraging K-means clustering and segmentation for image preprocessing on a dataset consisting of benign and malignant B-ALL cases, he compared the efficiency of three lightweight CNN models (EfficientNetB0, MobileNetV2, and NASNet Mobile) using the training and testing data. Eventually, segmented and original images were combined and fed as inputs through two channels to extract maximum feature space, which enhanced the models' accuracy. MobileNetV2 was selected for achieving 100\% accuracy and having the smallest size, making it suitable for implementation on mobile devices.

Talaat et al.~\cite{Talaat2023} exploited the attention mechanism to detect and classify leukemia cells. The A2M-LEUK algorithm involved image preprocessing, feature extraction using CNN, and an attention mechanism-based machine learning algorithm applied to the extracted features. The C-NMC 2019 ~\cite{Mourya2019} dataset was utilized, and classifiers such as SVM or a neural network were used in the proposed algorithm. After evaluating the precision, recall, accuracy, and specificity of the A2M-LEUK algorithm against KNN, SVM, Random Forest, and Naïve Bayes, the proposed model demonstrated superior performance, achieving nearly 100\% in all four metrics. However, the paper did not specify which classification model was used with A2M-LEUK to achieve that accuracy.

Yan~\cite{Yan2024} presented a study on the single-cell dataset C-NMC 2019~\cite{Mourya2019} to classify normal and cancerous white blood cells using three different models: YOLOv4, YOLOv8, and a CNN model. Data augmentation was applied to the CNN and YOLOv4 models. The CNN model, consisting of convolutional and max pooling layers, fully connected layers, and ReLU activation functions, achieved 93\% accuracy, while YOLOv4 and YOLOv8 both achieved accuracies above 95\%.

Devi et al.~\cite{Devi2024} utilized a combination of custom-designed and pretrained CNN architectures to detect ALL in the ALL image dataset~\cite{Ghaderzadeh2021} after applying augmentation. The custom-designed CNN was used to extract hierarchical features, while VGG-19 was used to extract high-level features. VGG-19 performed the classification task, and the proposed model achieved 97.85\% accuracy. On the other hand,~\cite{Khosrosereshki2017} applied image processing and the Fuzzy Rule-Based inference system to tackle the same topic.

Rahmani et al.~\cite{Rahmani2024} opted for the C-NMC 2019 dataset, where the data was preprocessed using methods such as grayscaling and masking, followed by feature extraction through transfer learning with models like VGG19, ResNet50, ResNet101, ResNet152, EfficientNetB3, DenseNet-121, and DenseNet-201. Feature selection was then applied using Random Forest, Genetic Algorithms, and the Binary Ant Colony Optimization metaheuristic algorithm. The classification was conducted through a multilayer perceptron, achieving an accuracy slightly above 90\%.

Kumar et al.~\cite{Kumar2020} contributed to the classification of different types of blood cancer in white blood cells, such as ALL and Multiple Myeloma. He applied preprocessing and augmentation methods to the data, followed by feature selection. The study used the SelectKBest class to select K specific features. The proposed model consisted of two blocks, each containing a convolutional layer and a max pooling layer, followed by fully connected layers and a classification layer. This architecture achieved 97.2\% accuracy.

Saikia et al.~\cite{Saikia2024} introduced VCaps-Net, a fine-tuned VGG16 model combined with a capsule network for ALL detection. Two datasets were used: ALL-IDB1 ~\cite{Genovese2023} and a private dataset. The proposed model integrates the powerful structure of VGG16 with a capsule network, which represents unit positions in images using vectors to maintain spatial relationships often lost due to max pooling. VCaps-Net achieved an accuracy of 98.64\%.

The ALL-IDB dataset was also used in a study by Alsaykhan et al.~\cite{Alsaykhan2024} to detect ALL using a hybrid algorithm. The approach combined support vector machine (SVM) and particle swarm optimization algorithms to optimize the results by selecting the best parameters to minimize errors. As a result, an accuracy of 97\% was achieved.

In~\cite{ABHISHEK2023}, Abhishek et al. classified different types of leukemia, including CLL, ALL, CML, and AML. He utilized the transfer learning approach, freezing the initial layers of pretrained CNNs as feature extractors (a process known as fine-tuning). The feature extractors used were ResNet152V2, MobileNet, DenseNet121, VGG16, InceptionV3, and Xception, which were trained on ImageNet~\cite{Russakovsky2015}. These extractors were then combined with classifiers such as Support Vector Machines, Random Forest, and new fully connected layers to improve classification performance. The accuracies for various combinations of classifiers ranged from 74\% to 84\%.

Vogado et al.~\cite{VOGADO2018} conducted a study using multiple datasets of different natures, focusing on multi-cell and single-cell images. CNNs were used for feature extraction from the original images, and SVM was applied for classification without prior image segmentation. The pre-trained models included AlexNet~\cite{Krizhevsky2012}, CaffeNet~\cite{Jia2014}, and VGG-f~\cite{Chatfield2014}. The feature vectors were then passed to the selected classifier for final predictions.
\section{ Comparison Analysis with Other Results}\label{sec:resultscomp}
In this section, we compare our image classification methods for Acute Lymphoblastic Leukemia (ALL) with existing approaches as shown in Table~\ref{tab:ALL_detection_comparison}. Our approach using YOLOv11s achieved an accuracy of 98.2\% on the ALL-IDB1 dataset, outperforming Yan’s YOLOv8 model, which reached 96\% on the C-NMC 2019 dataset. Additionally, our YOLOv8s model attained 98\%, while the Inception-ResNet-v2 model achieved 99.7\% accuracy, outperforming other custom CNN-based studies, such as Devi et al., which achieved 97.85\% on the ALL dataset. Our results also demonstrate a higher accuracy compared to the VCaps-Net model from Saikia et al., which obtained 98.64\% on ALL-IDB1, as well as the DenseNet-201 model by Rahmani et al., which achieved 90.55\% accuracy on the C-NMC 2019 dataset.

\begin{table}[ht]
    \centering
    \caption{Comparison of Different Approaches for Detecting Acute Lymphoblastic Leukemia (ALL)}
   \begin{tabular}{ |p{1.5cm}|p{1.8cm}|p{1.9cm}|p{1.6cm}|}
        \hline \hline
         & &  & \\
        \textbf{Study} & \textbf{Methodology} & \textbf{Accuracy} & \textbf{Dataset}\\
        & &  & \\
        \hline \hline
        Yan ~\cite{Yan2024}&YOLOv4&YOLOv4: 98\% &\\
        &YOLOv8&YOLOv8: 96\% &C-NMC 2019\\
        &CNN &CNN: 92\%&\\
        \hline
        Devi et al. \cite{Devi2024}& Custom + pretrained CNN & 97.85\% & ALL dataset \\
        \hline
        Rahmani et al. \cite{Rahmani2024}& DenseNet-201+ RF-GA-BACO & 90.55\% & C-NMC 2019 \\
        \hline
        Saikia et al. \cite{Saikia2024}& VCaps-Net & 98.64\% & ALL-IDB1 \\
        \hline
         Kumar et al. \cite{Kumar2020}& Custom CNN  & 97.2\% & Custom dataset \\
        \hline
        Abhishek et al. \cite{ABHISHEK2023}& SVM (VGG16 with LTCL fine tuned along with SVM)  & 84\% & Custom dataset \\
        \hline
        Alsaykhan et al. \cite{Alsaykhan2024}& SVM + PSO  & 97\% & ALL-IDB1 + ALL-IDB2 \\
        \hline
        Our study~\cite{awad2024}&YOLOv11n&YOLOv11n:97.3\%&\\
        &YOLOv11s&YOLOv11s: 98.2\%&ALL-IDB1\\
        &YOLOv8s&YOLOv8s: 98\%&+\\
        &ResNet50&ResNet50: 99\%&ALL dataset\\
        &Inception-ResNet-v2 &Inception-ResNet-v2: 99.7\%&\\
        \hline
    \end{tabular}
    \label{tab:ALL_detection_comparison}
\end{table}

\section{Conclusion}\label{sec:conclusion}
In conclusion, the integration of AI in the medical field is a massive step in the advancement of the health system and services provided to patients. In this research, computer vision techniques and several deep learning models were utilized to detect the absence or presence of Acute Lymphoblastic Leukemia in different stages. To achieve this goal, YOLOv11, YOLOv8, ResNet50, and Inception-ResNet-v2 were fine-tuned on multi-cell datasets to differentiate between healthy and cancerous white blood cells where this disease is usually found. In order to help the models learn diverse features, we collected images from ALL-IDB1 and ALL image datasets, then trained the models on the final integrated datasets. Consequently, the models exhibited high performances where the accuracies peaked at 97.3\%, 98.2\%, 98\%, 99\%, and 99.7\% for each of YOLOv11n, YOLO11vs, YOLO8vs, ResNet50, and Inception-ResNet-v2, respectively.

To the best of our knowledge, this is the first study to utilize the most recent and 11th version of the YOLO series. The comparison between YOLOv8 and YOLOv11 showed only slight differences in performance and proved that a model's complexity can play a noticeable role in its behavior. However, with data augmentation, other models such as ResNet50 and Inception-ResNet-v2 can perform better, especially as the network goes deeper.

For future work, we plan to collect more images from other datasets to boost the robustness of the models when encountering various features. This would assist the models in facing computer vision challenges pertaining to different settings and image preparations. We also aim to make our models suitable for deployment on different devices with consideration for the small hardware architectures.

\bibliographystyle{IEEEtran}
\bibliography{journalbloodcancerV14.bbl}

\end{document}